\documentclass[12pt]{iopart}

\usepackage{amsfonts, amssymb, mathrsfs, times, float, color, bm}

\usepackage{graphicx}
\usepackage{graphicx,epstopdf}
\usepackage{dcolumn}
\usepackage{bm}
\usepackage{makecell}

\newcommand{\nn}{\nonumber}

\usepackage{setspace}
\begin{document}

\title[Gravitational losses for the binary systems]{Gravitational losses for the binary systems induced by the next-to-leading spin-orbit coupling effects}

\author{Hao Zhang$^{\star,\dagger}$, ~~Wei Gao$^{\star,\dagger}$, ~~Guansheng He$^{\star}$, ~~Siming Liu$^{\star}$, ~~Huanyu Jia$^{\star}$, ~~Wenbin Lin$^{\star,\dagger}$}%
\address{$^{\star}$School of Mathematics and Physics, University of South China, Hengyang, 421001, China}
\address{$^{\dagger}$School of Physical Science and Technology, Southwest Jiaotong University, Chengdu, 610031, China}

\ead{lwb@usc.edu.cn (Wenbin Lin)}

%


\vspace{10pt}

\begin{abstract}
The orbital energy and momentum of the compact binary systems will loss due to gravitational radiation. Based on the mass and mass-current multipole moments of the binary system with the spin vector defined by Boh\'{e} et al. [Class. Quantum Grav. {\bf 30}, 075017 (2013)], we calculate the loss rates of energy, angular and linear momentum induced by the next-to-leading spin-orbit effects. For the case of circular orbit, the formulations for these losses are given in terms of the orbital frequency.
\end{abstract}


\vspace{1pc}

\section{introduction}

The compact binary systems such as black hole-black hole pairs, black hole-dense star (neutron star, white dwarf or others) pairs, or double dense stars,
are the best candidates for the gravitational-wave sources, which have been detected by LIGO/VIRGO~\cite{LIGO2016,Harry2010,Aasi2015,Acernese2015}, and and also will be detected by the space-based gravitational-wave detectors such as LISA, TianQin and Taiji in the near future~\cite{Acernese1997,Amaro-Seoane2017,Luo2016,Hu2017,Gong2021}.

Gravitational radiation will carry out the orbital energy, angular and linear momentum of the compact binary systems.
The analytic calculations for the gravitational losses can only be achieved via the post-Newtonian (PN) approximations, which have been extensively studied in literatures. The references in which the results are given for the first time are summarized in the following tables.

Table~\ref{T1} presents the references for the loss rates of the non-spinning binary system's energy $E$, angular momentum $\bm{J}$ and linear momentum $\bm{P}$ to the different PN orders including the tail contributions. The "$\cdot$" denotes the time derivative.
\begin{table*}[!htb]
\caption{References for the gravitational loss rates of the non-spinning binary\\
 systems in the Newtonian, 1PN, 2PN, 2.5PN, 3PN and 3.5PN approximation.}
\begin{tabular}{|c|c|c|c|}
\hline
gravitational loss rates  & circular orbit    & general orbit    \\
\hline\hline
{$\dot{E}_{N}$}      &\thead{Peters~\cite{PeterMathews1963}} &\thead{Peters~\cite{PeterMathews1963}}\\
\hline
{$\dot{E}_{PN}$}            &\thead{Will \& Wagoner~\cite{WagonerWill1976}}&\thead{Will \& Wagoner~\cite{WagonerWill1976}}\\
\hline
{$\dot{E}_{2PN}$}                             &\thead{ Blanchet, Damour \& Iyer~\cite{Blanchet1995}} &\thead{Will \& Wiseman~\cite{WillWiseman1996}}\\
\hline
{$\dot{E}_{2.5tail}$}            &\thead{Blanchet~\cite{Blanchet1996Energy}}  &\thead{Arun et al.~\cite{Arun2008}}\\
\hline
{$\dot{E}_{3PN}$}                             &\thead{Blanchet, Iyer \& Joguet~\cite{Blanchet2002}}  &\thead{Arun et al.~\cite{Arun2008}}\\
\hline
{$\dot{E}_{3.5tail}$}           &\thead{Blanchet~\cite{Blanchet1998Gravitational}}  &N/A\\
\hline
\hline
{$\dot{\bm{J}}_{N}$, $\dot{\bm{J}}_{PN}$}      &\thead{Junker \& Sch$\ddot{a}$fer~\cite{JunkerSchafer1992}} &\thead{Junker \& Sch$\ddot{a}$fer~\cite{JunkerSchafer1992}}\\
\hline
{$\dot{\bm{J}}_{2PN}$}            &\thead{Gopakumar \& Iyer~\cite{Gopakumart1997}} &\thead{Gopakumar \& Iyer~\cite{Gopakumart1997}}\\
\hline
{$\dot{\bm{J}}_{2.5tail}$, $\dot{\bm{J}}_{3PN}$}            &\thead{Arun et al.~\cite{Arun2009}}  &\thead{Arun et al.~\cite{Arun2009}}\\
\hline
\hline
{$\dot{\bm{P}}_{N}$}      &\thead{ Junker \& Sch$\ddot{a}$fer~\cite{JunkerSchafer1992}}  &\thead{ Junker \& Sch$\ddot{a}$fer~\cite{JunkerSchafer1992}}\\
\hline
{$\dot{\bm{P}}_{PN}$, $\dot{\bm{P}}_{2PN}$}            &\thead{Racine, Buonanno \& Kidder~\cite{Racine2009}}  &\thead{ Racine, Buonanno \& Kidder~\cite{Racine2009}}\\
\hline
{$\dot{\bm{P}}_{2.5tail}$}            &\thead{Kastha~\cite{Kastha2022}}  &\thead{Kastha~\cite{Kastha2022}}\\
\hline
\end{tabular}\label{T1}
\end{table*}

For the spinning binary systems, we need the spin vector and the spin supplementary condition (SSC) to characterize the systems' motion. In the literature, different definitions for the spin vector and different spin supplementary conditions have been employed. Although they describe the same physical phenomena, the formulations of the gravitational loss rates are dependent on the definitions for the spin vectors and the SSC. Table~\ref{T2} gives the reference for the loss rates of the spinning binary system's energy, angular momentum and linear momentum induced by the spin-orbit (SO) coupling based on the spin vector defined by Barker and O'Connell~\cite{Barker1974}. Kidder~\cite{Kidder1995} use this spin vector, and the SSC given by Pirani (called as P condition)~\cite{Pirani1956}, to calculate the lowest order of the SO contribution to gravitational radiation.

\begin{table*}[!htb]
\centering
\caption{Reference for the gravitational loss rates of the spinning binary systems\\
  with the spin vector defined by Barker and O'Connell and under the P condition.}
\begin{tabular}{|c|c|c|c|}
\hline
gravitational loss rates   & circular orbit    & general orbit    \\
\hline\hline
{$\dot{E}_{1.5SO}$, $\dot{\bm{J}}_{1.5SO}$, $\dot{\bm{P}}_{1.5SO}$}      &Kidder~\cite{Kidder1995}  & Kidder~\cite{Kidder1995}\\
\hline
\end{tabular}\label{T2}
\end{table*}

Faye, Blanchet and Buonanno~\cite{FayeBlanchetBuonanno2006} defines a new spin vector, and Blanchet, Buonanno and Faye calculate the 2.5PN order of SO contribution to the energy loss flux under the SSC given by Tulczyjew (called as T condition)~\cite{Tulczyjew1959}.
Table~\ref{T3} presents the formulations for the loss of the spinning binary system's energy, angular momentum and linear momentum duo to SO contributions, and spin-spin(SS) contributions base on this definition of the spin vector.
\begin{table*}[!htb]
\caption{References for the gravitational loss rates of the spinning binary systems\\
 base on the spin vector defined by Faye, Blanchet and Buonanno  and under\\
 the T condition.}\label{T3}
\begin{tabular}{|c|c|c|c|}
\hline
 gravitational loss rates    & circular orbit    & general orbit    \\
\hline\hline
$\dot{E}_{1.5SO}$    &\thead{Same as Kidder~\cite{Kidder1995}} &\thead{Same as Kidder~\cite{Kidder1995}}\\
\hline
{$\dot{E}_{2.5SO}$}      &\thead{Blanchet, Buonanno\\ \& Faye~\cite{BlanchetBuonanno2006}}  &\thead{Blanchet, Buonanno \\ \& Faye~\cite{BlanchetBuonanno2006}}\\
\hline\hline
\makecell{$\dot{\bm{J}}_{1.5SO}$  }            &\thead{Same as Kidder~\cite{Kidder1995}}  &\thead{Same as Kidder~\cite{Kidder1995}}\\
\hline
\makecell{$\dot{\bm{J}}_{2.5SO}$ } &\thead{\textbf{This work}} &\thead{ \textbf{This work}}\\
\hline\hline
\makecell{$\dot{\bm{P}}_{0.5SO}$ }            &\thead{Same as Kidder~\cite{Kidder1995}} &\thead{Same as Kidder~\cite{Kidder1995}}\\
\hline
\thead{\makecell{$\dot{E}_{2SS}$, $\dot{\bm{J}}_{2SS}$, $\dot{\bm{P}}_{1.5SO}$, $\dot{\bm{P}}_{2SS}$}} &\thead{Racine, Buonanno \& Kidder~\cite{Racine2009}} & \thead{Racine, Buonanno \& Kidder~\cite{Racine2009}}\\
\hline
\end{tabular}
\end{table*}
In 2013, Boh\'{e} et al. define a new spin vector~\cite{Bohe2013}, and with this spin vector, Marsat et al. calculate the spinning binary system's energy loss to the 3.5PN SO and 4PN tail-induced term contributions for the case of circular orbit under the T condition~\cite{Marsat2014}. In this work, we extend the latter's work to the case of general orbit, and also calculate the loss rates of the angular and linear momentum. Table~\ref{T4} presents the references for the gravitational loss rates of the spinning binary systems due to the SO and SS contributions base on this definition of the spin vector.

\begin{table*}[!htb]
\centering
\caption{References for the gravitational loss rates of the spinning binary systems based \\
on the spin vector defined by Boh\'{e} et al. and under the T condition.}\label{T4}
\begin{tabular}{|c|c|c|c|}
\hline
gravitational loss rates    & circular orbit    & general orbit    \\
\hline\hline
{$\dot{E}_{1.5SO}$}      &\thead{Same as Kidder~\cite{Kidder1995}}  &\thead{Same as Kidder~\cite{Kidder1995}}\\
\hline
{$\dot{E}_{2.5SO}$}      &\thead{Marsat et al.~\cite{Marsat2014}}  &\thead{\textbf{This work}}\\
\hline
{$\dot{E}_{3SOtail}$, $\dot{E}_{3.5SO}$, $\dot{E}_{4SOtail}$}      &\thead{Marsat et al.~\cite{Marsat2014}}  &N/A\\
\hline
{$\dot{E}_{2SS}$, $\dot{E}_{3SS}$}      &\thead{Boh\'{e} et al.~\cite{Bohe2015}}  &\thead{Boh\'{e} et al.~\cite{Bohe2015}}\\
\hline\hline
{$\dot{\bm{J}}_{1.5SO}$}            &\thead{Same as Kidder~\cite{Kidder1995}}  &\thead{Same as Kidder~\cite{Kidder1995}}\\
\hline
{$\dot{\bm{J}}_{2.5SO}$  } &\textbf{This work} & \textbf{This work}\\
\hline\hline
{$\dot{\bm{P}}_{0.5SO}$}            &\thead{Same as Kidder~\cite{Kidder1995}}  &\thead{Same as Kidder~\cite{Kidder1995}}\\
\hline
{$\dot{\bm{P}}_{1.5SO}$} &\textbf{This work} & \textbf{This work}\\
\hline
\end{tabular}
\end{table*}

The rest of this paper is organized as follows. Section~\ref{sec2} introduces the 2.5PN acceleration for the relative motion of the binary systems, which will be used in later derivations. In Section~\ref{sec3} we give the formulas for calculating gravitational losses, Section~\ref{sec4} we derive the gravitational losses induced by the next-to-leading spin-orbit coupling effects. As a comparison, we also present the angular momentum loss in terms of the spin vector defined by Faye, Blanchet and Buonanno in Appendix B. Section~\ref{sec5} gives these loss rates for circular orbit. Summary is given in Section~\ref{sec6}.
In this article, small Greek alphabet represent 0, 1, 2, 3, and small letters represent 1, 2, 3.

\section{The motion for the binary system in the 2.5PN approximation}\label{sec2}
The problem of the motion of a rotating object in a gravitational field is first investigated by Mathisson~\cite{Mathisson1937}. Subsequently, the equations-of-motion of the spinning body in general relativity are developed by Papapetrou, Tulczyjew, and Dixon
~\cite{Papapetrou1951,Tulczyjew1959,Dixon1964}, which are called as the MPTD equations:
\begin{eqnarray}
&&\frac{Dp^{\mu}}{D\tau}=-\frac{1}{2}R_{~~\nu\kappa\lambda}^{\mu}u^{\nu}S^{\kappa\lambda}~,\label{MPTDp}
\end{eqnarray}
\begin{eqnarray}
&&\frac{D{S}^{\mu\nu}}{D\tau}=p^{\mu}u^{\nu}-p^{\nu}u^{\mu}~.\label{MPTDS}
\end{eqnarray}

where, $\frac{D}{D\tau}$  denotes absolute differentiation with respect
to the proper time $\tau$ of the particle. $S^{\mu\nu}$ represents the spin tensor of the object: $S_{A}^{\mu\nu}=2\int_{A}(X^{[\mu}-X_{A}^{[\mu})\tau^{\nu]0}d^{3}x$,  $X^{\mu}$ represents the center of mass of body A, $\tau^{\nu0}$ represents the stress-energy tensor. $u^{\mu}$ and $p^{\mu}$ are four velocity and four momentum, respectively. $R_{~~\nu\kappa\lambda}^{\mu}$ is curvature tensor. Since Eqs.\,(\ref{MPTDp}) and (\ref{MPTDS}) only have 7 independent equations but the unknown number is 10, we need the spin supplementary condition, which represents the definition of different center-of-mass world line. At the same time, there are also different ways to define the spin vector.

Kidder~\cite{Kidder1995} use the spin vector defined by Barker and O'Connell:
\begin{eqnarray}
\bar{S}_A^i=\frac{1}{2}\epsilon^{ijk}\bar{S}_A^{jk}
\end{eqnarray}
where $\epsilon^{ijk}$ is the three-dimensional antisymmetric tensor of Levi-Civita, A denotes body A, and the P condition: $S^{\mu\nu}u_{\mu}=0$, gives the 1.5PN order of SO contribution to the the equation of motion. Blanchet, Buonanno and Faye~\cite{FayeBlanchetBuonanno2006} defines the spin vector $\tilde{S}_{\sigma}$ satisfying:
\begin{eqnarray}
\tilde{S}_A^{\mu\nu}=-\frac{1}{\sqrt{g_A}}\epsilon^{\mu\nu\rho\sigma}\frac{p_{\rho}^A}{m}\tilde{S}_{\sigma}^A\label{Sd06}
\end{eqnarray}
where $\epsilon^{\mu\nu\rho\sigma}$ is the four-dimensional antisymmetric tensor of Levi-Civita, m denotes the mass of the body, $g_{A}$ denotes the determinant of the metric at point A, and use the T condition: $S^{\mu\nu}p_{\mu}\!=\!0$, gives the 2.5PN order of SO contribution to the the equations of motion and the 2PN order precession equations of spin vector. Boh\'{e} et al.~\cite{Bohe2013} define a new spin vector such that:
\begin{eqnarray}
S_{A}^{i}=H_{A}^{ij}\tilde{S}_{j}^{A}
\end{eqnarray}
with $H_{A}^{ij}$ being the unique symmetric and positive-definite square root of the matrix $G^{ij}=g^{ij}\!-\!2g^{0(i}u^{j)}/u^0\!+\!g^{00}u^iu^j/(u^0)^2$ at the particle's position
A, and derive the 3.5PN dynamic equation of the center-of-mass and the 3PN precession equations of the spin vector under the T condition. The advantages for this new definition of the spin vector can be found in their work.In this work, we use the 2.5PN dynamic equation of the center-of-mass and the 1PN precession equations of the spin vector given by Boh\'{e} et al.~\cite{Bohe2013}, and the mass moment given by Marsat et al.~\cite{Marsat2014}, to calculate the gravitational loss of the spinning binary systems  induced by the next-to-leading SO coupling effect.

We assume the spinning compact binary has masses $M_1$ and $M_2$. The position vectors of the bodies are $\bm{X}_1$ and $\bm{X}_2$, and the corresponding velocities are $\bm{V}_1$ and $\bm{V}_2$, the spins of the two bodies are $\bm{S}_1$ and $\bm{S}_2$. The precession equations of the total spin vector $\bm{S}=\bm{S}_1+\bm{S}_2$ can be written as~\cite{Bohe2013}:
\begin{eqnarray}
&&\frac{d\bm{S}}{dt}=\frac{GM}{c^2R^2}\eta\Big\{\mathbf{n}\Big[-\frac{7}{2}(\bm{V}\cdot\bm{S})-\frac{3}{2}\frac{\delta M}{M}(\bm{V}\cdot\bm{\Delta})\Big]\nn\\
&&\hskip 1.0cm +\, V\bm{\lambda}\Big[\frac{7}{2}(\bm{n}\cdot\bm{S})+\frac{3}{2}\frac{\delta M}{M}(\bm{n}\cdot\bm{\Delta})\Big]\Big\}~,\label{ds}
\end{eqnarray}
\begin{eqnarray}
&& \frac{d\bm{\Delta}}{dt}=\frac{GM}{c^2R^2}\Big\{\mathbf{n}\Big[-\frac{3}{2}\frac{\delta M}{M}(\bm{V}\cdot\bm{S})-\Big(\frac{3}{2}-\frac{5}{2}\eta\!\Big)(\bm{V}\cdot\bm{\Delta})\Big]\nn\\
&&\hskip 1.0cm +V\bm{\lambda}\Big[\frac{3}{2}\frac{\delta M}{M}(\bm{n}\cdot\bm{S})+\Big(\frac{3}{2}-\frac{5}{2}\eta\Big)(\bm{n}\cdot\bm{\Delta})\Big ]\Big\}~.\label{dd}
\end{eqnarray}
where $M\!=\! M_1+M_2$ denotes the total mass of the system, $\delta M=M_1-M_2$ and $\eta=M_1M_2/M^2$. $\bm{\Delta}=M\Big(\frac{\bm{S}_2}{M_2}-\frac{\bm{S}_1}{M_1}\Big)$.
The unit vector $\bm{n}\!\equiv\! \bm{R}/R$ with $\bm{R}\!=\!\bm{X}_1 \!-\! \bm{X}_2$ being the separation vector between the two bodies and $R\!\equiv\! |\bm{R}|$. The unit vector $\bm{\lambda}\!\equiv\!\bm{V}/V$ with $\bm{V}\!=\!\bm{V}_1\!-\!\bm{V}_2$ being the relative velocity of the bodies and $V\!=\!|\bm{V}|$. For later use, we use $\bm{l}\!=\!\bm{n}\!\times\!\bm{\bm{\lambda}}$ to denote the unit vector for the angular momentum.
We also need the 2.5PN acceleration of the binary system's relative motion, which can be written as~\cite{Kidder1995,Bohe2013}
\begin{eqnarray}
\frac{d\bm{V}}{dt}=\bm{A_{N}}+\bm{A_{PN}}+\bm{A_{1.5SO}}+\bm{A_{2PN}}+\bm{A_{2.5SO}}~,\label{Acceleration-2.5PN}
\end{eqnarray}
where
\begin{eqnarray}
&&\hskip -0.75cm\bm{A_{N}}=-\frac{GM}{R^2}\bm{n}~,\label{A-N}
\end{eqnarray}
\begin{eqnarray}
&&\hskip -1.75cm\bm{A_{PN}}=-\frac{GM}{c^2R^2}\Big\{\bm{n}\Big[(1+3\eta)V^2-(4+2\eta)\frac{GM}{R}-\frac{3}{2}\eta\dot{R}^2\Big]
\,-\,2(2-\eta)\dot{R}V\bm{\lambda}\Big\}~,\label{A-PN}
\end{eqnarray}
\begin{eqnarray}
&&\hskip -0.75cm\bm{A_{1.5SO}}=\frac{G}{c^3R^3}\Big\{6\bm{n}\Big[2(\bm{n}\times\bm{V})\cdot\bm{S}+\frac{\delta m}{m}(\bm{n}\times\bm{V})\cdot\bm{\Delta}\Big]
-\Big[7(\bm{V}\times\bm{S})\nn \\
&&\hskip -0.75cm \hskip 1.45cm+3\frac{\delta m}{m}(\bm{V}\times\bm{\Delta})\Big]+3\dot{R}\Big(3\bm{n}\times\bm{S}+\frac{\delta m}{m}\bm{n}\times\bm{\Delta}\Big)\Big\}~,\label{A-1.5SO}
\end{eqnarray}
\begin{eqnarray}
&&\hskip -1.75cm\bm{A_{2PN}}=-\frac{GM}{c^4R^2}\Big\{\bm{n}\Big[\frac{3}{4}(12+29\eta)\frac{(GM)^2}{R^2}+\eta(3-4\eta)V^4
+\frac{15}{8}\eta(1-3\eta)\dot{R}^4\nn\\
&&\hskip -1.75cm \hskip 1.45cm-\frac{3}{2}\eta(3-4\eta)V^2\dot{R}^2 -\frac{1}{2}\eta(13-4\eta)\frac{GM}{R}V^2-(2+25\eta+2\eta^2)\frac{GM}{R}\dot{R}^2\Big]\nn\\
&&\hskip -1.75cm \hskip  1.45cm
-\frac{1}{2}\dot{R}V\bm{\lambda}\Big[\eta(15+4\eta)V^2-(4+41\eta+8\eta^2)\frac{GM}{R}-3\eta(3+2\eta)\dot{R}^2\Big]\Big\}~,\label{A-2PN}
\end{eqnarray}
\begin{eqnarray}
&&\hskip -0.75cm \bm{A_{2.5SO}}=\frac{G}{c^5R^3}\Big\{\Big[(24+19\eta)\frac{GM}{R}+\frac{3}{2}(1+10\eta)\dot{R}^2-14\eta V^2\Big](\bm{V} \times\bm{S})\nn\\
&&\hskip -0.75cm \hskip 1.45cm-\Big[(28+29\eta)\frac{GM}{R}+\frac{45}{2}\eta\dot{R}^2+\frac{3}{2}(1-15\eta)V^2\Big]\dot{R}(\bm{n}\times\bm{S})\nn\\
&&\hskip -0.75cm \hskip 1.45cm+\Big[\frac{1}{2}(24+19\eta)\frac{GM}{R}+\frac{3}{2}(1+6\eta)\dot{R}^2-7\eta V^2\Big]\frac{\delta M}{M}(\bm{V}\times\bm{\Delta})\nn\\
&&\hskip -0.75cm \hskip 1.45cm-\Big[\frac{1}{2}(24+31\eta)\frac{GM}{R}+15\eta\dot{R}^2+\frac{3}{2}(1-8\eta)V^2\Big]\frac{\delta M}{M}\dot{R}^2(\bm{n}\times\bm{\Delta})\nn\\
&&\hskip -0.75cm \hskip 1.45cm-\bm{n}\Big[(44+33\eta)\frac{GM}{R}+30\eta\dot{R}^2-24\eta V^2\Big](\bm{n} \times\bm{V})\cdot \bm{S}\nn\\
&&\hskip -0.75cm \hskip 1.45cm-\bm{n}\Big[\frac{1}{2}(48+37\eta)\frac{GM}{R}+15\eta\dot{R}^2-12\eta V^2\Big]\frac{\delta M}{M}(\bm{n} \times\bm{V})\cdot \bm{\Delta}\nn\\
&&\hskip -0.75cm \hskip 1.45cm-\Big[\frac{21}{2}(1-\eta)(\bm{n} \times\bm{V})\cdot \bm{S}+\frac{3}{2}(3-4\eta)\frac{\delta M}{M}(\bm{n} \times\bm{V})\cdot\bm{\Delta}\Big]\dot{R}V\bm{\lambda}\Big\}~.\label{A-2.5SO}
\end{eqnarray}
here $\dot{R}=\bm{n}\!\cdot\!\bm{V}$.
\section{Formulas for calculating gravitational losses}\label{sec3}
The isolated system's energy, angular momentum, and linear momentum losses due to the gravitational-wave radiation can be written in terms of the symmetric and tracefree (STF)-multipole moments as follows~\cite{Thorne1980}:
{\small\begin{eqnarray}
&& \frac{dE}{dt}=-\frac{G}{c^5}\sum_{l=2}^\infty\Big\{\Big(\frac{1}{c}\Big)^{2(l-2)}\frac{(l+1)(l+2)}{l(l-1)l!(2l+1)!!}
    \mathop{\mathcal{I}_{A_l}}\limits_{}^{(l+1)}\mathop{\mathcal{I}_{A_l}}\limits_{}^{(l+1)}\nn\\
&&\hskip 1.0cm +\,\Big(\frac{1}{c}\Big)^{2(l-1)}\frac{4l(l+2)}{(l-1)(l+1)!(2l+1)!!}
\mathop{\mathcal{J}_{A_l}}\limits_{}^{(l+1)}\mathop{\mathcal{J}_{A_l}}\limits_{}^{(l+1)}\Big\}~,\label{identity-Loss-E}
\end{eqnarray}}
{\small\begin{eqnarray}
&& \frac{dJ_j}{dt}=-\frac{G}{c^5}\sum_{l=2}^\infty\Big\{\Big(\frac{1}{c}\Big)^{2(l-2)}\!\!\!\!\!\frac{(l+1)(l+2)}{(l-1)l!(2l+1)!!}
\epsilon_{jpq}\mathop{\mathcal{I}_{pA_{l-1}}}\limits_{}^{l}\mathop{\mathcal{I}_{qA_{l-1}}}\limits_{}^{l+1}\nn\\
&&\hskip 1.0cm +\,\Big(\frac{1}{c}\Big)^{2(l-1)}\!\!\!\!\!\!\!\frac{4l^2(l+2)}{(l-1)(l+1)!(2l+1)!!}\epsilon_{jpq}
\mathop{\mathcal{J}_{pA_{l-1}}}\limits_{}^{l}\mathop{\mathcal{J}_{qA_{l-1}}}\limits_{}^{l+1}\Big\}~,\label{identity-Loss-J}
\end{eqnarray}}
{\small\begin{eqnarray}
&& \frac{dP_j}{dt}=-\frac{G}{c^7}\sum_{l=2}^\infty\Big\{\Big(\frac{1}{c}\Big)^{2(l-2)}\!\!\frac{2(l+2)(l+3)}{l(l+1)!(2l+3)!!}
\mathop{\mathcal{I}_{jA_l}}\limits_{}^{(l+2)}\mathop{\mathcal{I}_{A_l}}\limits_{}^{(l+1)}\!\!\!\nn\\
&&\hskip 1.0cm+\!\Big(\frac{1}{c}\Big)^{2(l-1)}\frac{8(l+3)}{(l+1)!(2l+3)!!}
\mathop{\mathcal{J}_{jA_l}}\limits_{}^{(l+2)}\mathop{\mathcal{J}_{jA_l}}\limits_{}^{(l+1)}\nn\\
&&\hskip 1.0cm+\,\Big(\frac{1}{c}\Big)^{2(l-2)}\frac{8(l+2)}{(l-1)(l+1)!(2l+1)!!}
\epsilon_{jpq}\mathop{\mathcal{I}_{pA_{l-1}}}\limits_{}^{(l+1)}\mathop{\mathcal{J}_{qA_{l-1}}}\limits_{}^{(l+1)}\Big\}~.\label{identity-Loss-P}
\end{eqnarray}}

where $E$, $J_j$ and $P_j$ are the orbital energy, angular momentum and linear one of the system.
 $\mathcal{I}_{A_l}$ and $\mathcal{J}_{A_l}$ are the STF radiative mass and current multipole moments, respectively. $A_l$ denotes a multi-index of length $l$, i.e. $A_l=a_1a_2...a_l$, where the $a$'s take the indies of $1,2,3$.
$\mathop{\mathcal{I}}\limits_{}^{(l)}\!\equiv\!d^l\mathcal{I}/dt^l$ and $\mathop{\mathcal{J}}\limits_{}^{~(l)}\!\equiv\!d^l\mathcal{J}/dt^l$.

Following the calculation method given in Ref.~\cite{JunkerSchafer1992}, we take all possible values for the parameters $l$ in Eqs. (\ref{identity-Loss-E})-(\ref{identity-Loss-P}) to ensure the accuracy of $\frac{dE}{dt}$ and $\frac{dJ_j}{dt}$ to the 2.5PN order, and that of $\frac{dP_j}{dt}$ to the 1.5PN order, as follows

\begin{eqnarray}
&& \frac{dE}{dt}= -\frac{G}{c^5}\Big[\frac{1}{5}\mathop{{\mathcal{I}}_{kl}}\limits_{}^{(3)}\mathop{{\mathcal{I}}_{kl}}\limits_{}^{(3)}
+\frac{1}{c^2}\frac{16}{45}\mathop{\mathcal{J}_{kl}}\limits_{}^{(3)}\mathop{\mathcal{J}_{kl}}\limits_{}^{(3)}
+\frac{1}{c^2}\frac{1}{189}\mathop{\mathcal{I}_{klm}}\limits_{}^{(4)}\mathop{\mathcal{I}_{klm}}\limits_{}^{(4)}\nn\\
&&\hskip1.0cm+\frac{1}{c^4}\frac{1}{84}\mathop{\mathcal{J}_{klm}}\limits_{}^{(4)}\mathop{\mathcal{J}_{klm}}\limits_{}^{(4)}+\frac{1}{c^4}\frac{1}{9072}\mathop{\mathcal{I}_{klmn}}\limits_{}^{(5)}\mathop{\mathcal{I}_{klmn}}\limits_{}^{(5)}\Big]~,\label{identity-E}
\end{eqnarray}

\begin{eqnarray}
&& \frac{dJ_j}{dt}= -\frac{G}{c^5}\epsilon_{jpq}\Big[\frac{2}{5}\mathop{\mathcal{I}_{pk}}\limits_{}^{(2)}\mathop{\mathcal{I}_{qk}}\limits_{}^{(3)}
+\frac{1}{c^2}\frac{32}{45}\mathop{{\mathcal{J}}_{pk}}\limits_{}^{(2)}\mathop{{\mathcal{J}}_{qk}}\limits_{}^{(3)}
+\frac{1}{c^2}\frac{1}{63}\mathop{{\mathcal{I}}_{pkl}}\limits_{}^{(3)}\mathop{{\mathcal{I}}_{qkl}}\limits_{}^{(4)}\nn\\
&&\hskip1.0cm+\frac{1}{c^4}\frac{1}{28}\mathop{{\mathcal{J}}_{pkl}}\limits_{}^{(3)}\mathop{{\mathcal{J}}_{qkl}}\limits_{}^{(4)}
+\frac{1}{c^4}\frac{1}{2268}\mathop{{\mathcal{I}}_{pklm}}\limits_{}^{(4)}\mathop{{\mathcal{I}}_{qklm}}\limits_{}^{(4)}\Big],\label{identity-J}
\end{eqnarray}

\begin{eqnarray}
&& \frac{dP_j}{dt}= -\frac{G}{c^7}\Big[\frac{2}{63}\mathop{{\mathcal{I}}_{jkl}}\limits_{}^{(4)}\mathop{{\mathcal{I}}_{kl}
}\limits_{}^{(3)}+\frac{1}{c^2}\frac{4}{63}\mathop{{\mathcal{J}}_{jkl}}\limits_{}^{(4)}\mathop{{\mathcal{J}}_{kl}}\limits_{}^{(3)}
+\frac{16}{45}\epsilon_{jpq}\mathop{{\mathcal{I}}_{pk}}\limits_{}^{(3)}\mathop{{\mathcal{J}}_{qk}}\limits_{}^{(3)}\nn\\
&&\hskip1.0cm+\frac{1}{c^2}\frac{1}{1134}\mathop{{\mathcal{I}}_{jklm}}\limits_{}^{(5)}\mathop{{\mathcal{I}}_{klm}}\limits_{}^{(4)}
+\frac{1}{c^2}\frac{1}{126}\epsilon_{jpq}\mathop{{\mathcal{I}}_{pkl}}\limits_{}^{(4)}\mathop{{\mathcal{J}}_{qkl}}\limits_{}^{(4)}\Big]~.\label{identity-P}
\end{eqnarray}
The non-spin mass and mass-current multipole moments for the binary system can be written as~\cite{WillWiseman1996,Damour2020}
\begin{eqnarray}
&&\hskip-2.0cm\mathop{{\mathcal{I}_{ij}}}\limits_{NS}=\mu R^2\Big\{1+\frac{1}{c^2}\Big[\frac{29}{42}(1-3\eta)V^2-\frac{1}{7}(5-8\eta)\frac{GM}{R}\Big]\nn\\
&&\hskip-1.0cm +\frac{1}{c^4}\Big[\frac{1}{756}(2021-5947\eta-4883\eta^2)\frac{GM}{R}V^2-\frac{1}{252}(355+1906\eta-337\eta^2)\frac{(GM)^2}{R^2}\nn\\
&&\hskip-1.0cm -\frac{1}{756}(131-907\eta+1273\eta^2)\frac{GM}{R}\dot{R}^2
+\frac{1}{504}(253-1835\eta+3545\eta^2)V^4\Big]\Big\}n_{<ij>}\nn\\
&&\hskip-1.0cm+\mu R^2\Big\{\frac{11}{21c^2}(1-3\eta)+\frac{1}{c^4}\Big[\frac{1}{189}(742-335\eta-985\eta^2)\frac{GM}{R}\nn\\
&&\hskip-1.0cm+\frac{1}{126}(41-337\eta+733\eta^2)V^2+\frac{5}{63}(1-5\eta+5\eta^2)\dot{R}^2\Big]\Big\}V_{<ij>}\nn\\
&&\hskip-1.0cm-\mu\dot{R}R^2\Big\{\frac{4}{7c^2}(1-3\eta)+\frac{2}{c^4}\Big[\frac{1}{63}(13-101\eta+209\eta^2)V^2\nn\\
&&\hskip-1.0cm+\frac{1}{756}(1085-4057\eta-1463\eta^2)\frac{GM}{R}\Big]\Big\}n_{<i}V_{j>}~,\label{Iij}
\end{eqnarray}
\begin{eqnarray}
&&\hskip-2.0cm\mathop{\mathcal{J}_{ij}}\limits_{NS}=\mu R^2\frac{\delta M}{M}\Big\{1+\frac{1}{c^2}\Big[\frac{3}{14}(9+10\eta)\frac{GM}{R}
+\frac{1}{28}(13-68\eta)V^2\Big]\Big\}\epsilon_{pq<i}n_{j>}n_{p}V_{q}\nn \\
&&\hskip -1.0cm-\mu\frac{\delta M}{M}\frac{1}{c^2}\frac{5}{28}(1-2\eta)\dot{R}R^2{\epsilon_{pq<i}V_{j>}n_{p}V_{q}}~,\label{Jij}
\end{eqnarray}
\begin{eqnarray}
&&\hskip-2.0cm\mathop{\mathcal{I}_{ijk}}\limits_{NS}=\mu R^3\frac{\delta M}{M}\Big\{-1+\frac{1}{c^2}\Big[\frac{1}{6}(5-13\eta)\frac{GM}{R}-\frac{1}{6}(5-19\eta)V^2\Big]\Big\}{n_{<ijk>}}\nn\\
&&\hskip-1.0cm+\mu\frac{\delta M}{M}\frac{1}{c^2}(1-2\eta)\dot{R}R^3{n_{<ij}V_{k>}}-\mu\frac{\delta M}{M}\frac{1}{c^2}(1-2\eta)R^3{n_{<i}V_{jk>}}~,\label{Iijk}
\end{eqnarray}
\begin{eqnarray}
&&\hskip-2.0cm\mathop{\mathcal{J}_{ijk}}\limits_{NS}=\mu(1-3\eta)R^3{\epsilon_{pq<i}n_{jk>}n_{p}V_{q}}~,\label{Jijk}
\end{eqnarray}
\begin{eqnarray}
&&\hskip-2.0cm\mathop{\mathcal{I}_{ijkl}}\limits_{NS}=\mu(1-3\eta)R^4{n_{<ijkl>}}~.\label{Iijkl}
\end{eqnarray}
where ${n_{i}}$ and ${V_{i}}$ denote the $i$ component of $\bm{n}$ and $\bm{V}$. $\mu=\eta M$ is the reduce mass of the binary system. $n_{<ij>}=n_{<i}n_{j>}$, etc. denote the STF section of the tensors, as in Ref.~\cite{JunkerSchafer1992}.

The spin mass and mass-current multipole moments for the binary system with the spin vector defined by Boh\'{e} et al. can be written as~\cite{Marsat2014}
\begin{eqnarray}
&&\hskip-2.5cm\mathop{\mathcal{I}_{ij}}\limits_{S}=\frac{R\eta}{c^3}\Big\{\!\frac{8}{3}(\bm{V}\times\bm{S})_{<i}n_{j>}
+\frac{8}{3}\frac{\delta M}{M}(\bm{V}\times\bm{\Delta})_{<i}n_{j>}-\frac{4}{3}(\bm{n}\times\bm{S})_{<i}V_{j>}
\nn\\
&&\hskip -1.5cm-\frac{4}{3}\frac{\delta M}{M}(\bm{n}\times\bm{\Delta})_{<i}V_{j>}\!\Big\}\!+\!\frac{R\eta}{c^5}\Big\{\!\Big[\!\frac{1}{21}(33+125\eta)\frac{GM}{R}\!+\!\frac{26}{21}(1-3\eta)V^2\!\Big](\!\bm{V}\times\bm{S}\!)_{<i}n_{j>}\nn\\
&&\hskip-1.5cm+\Big[\frac{1}{3}(1+16\eta)\frac{GM}{R}+\frac{1}{21}(26-116\eta)V^2\Big]\frac{\delta M}{M}(\bm{V}\times\bm{\Delta})_{<i}n_{j>}\nn\\
&&\hskip -1.5cm+\Big[-\frac{1}{3}(22+10\eta)\frac{GM}{R}-\frac{4}{21}(1-3\eta)V^2\Big](\bm{n}\times\bm{S})_{<i}V_{j>}\nn\\
&&\hskip -1.5cm+\Big[-\frac{1}{21}(56+34\eta)\frac{GM}{R}+\frac{1}{21}(-4+36\eta)V^2\Big]\frac{\delta M}{M}(\bm{n}\times\bm{\Delta})_{<i}V_{j>}\nn\\
&&\hskip -1.5cm-\frac{4}{21}(1-3\eta)\dot{R}(\bm{V}\times\bm{S})_{<i}V_{j>}-\frac{4}{21}(1-5\eta)\dot{R}\frac{\delta M}{M}(\bm{V}\times\bm{\Delta})_{<i}V_{j>}\nn\\
&&\hskip -1.5cm+\Big[\frac{3}{7}(1-3\eta)(\bm{n}\times\bm{V})\cdot \bm{S}+\frac{1}{21}(9-40\eta)\frac{\delta M}{M}(\bm{n}\times\bm{V})\cdot\bm{\Delta}\Big]V_{<i}V_{j>}\nn\\
&&\hskip -1.5cm+\frac{GM}{R}\Big[\frac{38}{21}(1+12\eta)(\bm{n}\times\bm{V})\cdot \bm{S}+\frac{2}{21}(24-13\eta)\frac{\delta M}{M}(\bm{n}\times\bm{V})\cdot\bm{\Delta}\Big] n_{<i}n_{j>}\nn\\
&&\hskip -1.5cm+\frac{GM}{R}\Big[\frac{1}{21}(17+61\eta)\dot{R}(\bm{n}\times\bm{S})_{<i}n_{j>}+\frac{1}{21}(21+34\eta)\dot{R}\frac{\delta M}{M}(\bm{n}\times\bm{\Delta})_{<i}n_{j>}\Big] \nn\\
&&\hskip -1.5cm+\frac{GM}{R}\Big[-\frac{1}{3}(6-10\eta)(\bm{n}\cdot\bm{S})-\frac{2}{3}(3-2\eta)\frac{\delta M}{M}(\bm{n}\cdot\bm{\Delta})\Big](\bm{n}\times\bm{V})_{<i}n_{j>} \Big\}~,\label{IijS}
\end{eqnarray}
\begin{eqnarray}
&&\hskip -2.5cm\mathop{\mathcal{J}_{ij}}\limits_{S}=-\frac{3}{2}\frac{R\eta}{c}n_{<i}\Delta_{j>}+\frac{R\eta}{c^3}
\Big\{\Big[-\frac{2}{7}\frac{\delta M}{M}V^2+\frac{10}{7}\frac{\delta M}{M}\frac{GM}{R}\Big]n_{<i}S_{j>}+\Big[-\frac{1}{28}(29-143\eta)V^2\nn \\
&&\hskip -1.5cm+\frac{1}{28}(61-71)\frac{GM}{R}\Big]n_{<i}\Delta_{j>}+\frac{1}{28}\Big[33\frac{\delta M}{M}(\bm{V}\cdot\bm{S})+(33-155\eta)(\bm{V}\cdot\bm{\Delta})\Big]n_{<i}V_{j>}\nn \\
&&\hskip -1.5cm+\frac{1}{7}\dot{R}\Big[3\frac{\delta M}{M}V_{<i}S_{j>}+(3-16\eta)V_{<i}\Delta_{j>}\Big]-\frac{1}{14}\Big[(11-47\eta)(\bm{n}\cdot\bm{\Delta})\nn \\
&&\hskip -1.5cm+11\frac{\delta M}{M}(\bm{n}\cdot\bm{S})\Big]V_{<i}V_{j>}\!-\!\frac{1}{14}\frac{GM}{R}\!\Big[\!29\frac{\delta M}{M}(\bm{n}\cdot\bm{S})\!+\!(8-31\eta)(\bm{n}\cdot\bm{\Delta})\!\Big]n_{<i}n_{j>}\!\Big\}\!,\!\label{JijS}
\end{eqnarray}
\begin{eqnarray}
&&\hskip -2.5cm\mathop{\mathcal{I}_{ijk}}\limits_{S}=\frac{R^2\eta}{c^3}\Big\{\!-\!\frac{9}{2}\frac{\delta M}{M}(\bm{V}\times\bm{S})_{<i}n_{j}n_{k>}-
\frac{3}{2}(3-11\eta)(\bm{V}\times\bm{\Delta})_{<i}n_{j}n_{k>\!}\nn \\
&&\hskip -1.cm+3\frac{\delta M}{M}(\bm{n}\times\bm{S})_{<i}n_{j}V_{k>}+3(1-3\eta)(\bm{n}\times\bm{\Delta})_{<i}n_{j}V_{k>}\Big\}~,\label{IijkS}
\end{eqnarray}
\begin{eqnarray}
\hskip -2.5cm\mathop{\mathcal{J}_{ijk}}\limits_{S}=\frac{R^2\eta}{c}\Big[2n_{<ij}S_{k>}}+2\frac{\delta M}{M}{n_{<ij}\Delta_{k>}\Big]~.\label{JijkS}
\end{eqnarray}
The STF tensors used in our derivations are enclosed in the Appendix A.
\section{Gravitational losses induced by the next-to-leading spin-orbit coupling effects}\label{sec4}
Substituting Eqs.~(\ref{Iij})-(\ref{JijkS}) into Eqs.~(\ref{identity-E})-(\ref{identity-P}), making use of Eqs.~(\ref{A-N})-(\ref{A-2.5SO}) to calculate the time derivative of velocity, and keeping the results to include the contributions of the next-to-leading spin-orbit coupling effects, we can obtain the loss rates of the orbital energy, angular momentum and linear momentum as follows

\begin{eqnarray}
&& \frac{dE}{dt}=\dot{E}_{N}+\dot{E}_{PN}+\dot{E}_{1.5SO}+\dot{E}_{2PN}+\dot{E}_{2.5SO}~,\label{identity-E1}
\\
&& \frac{d\bm{J}}{dt}=\dot{\bm{J}}_{N}+\dot{\bm{J}}_{PN}+\dot{\bm{J}}_{1.5SO}+\dot{\bm{J}}_{2PN}+\dot{\bm{J}}_{2.5SO}~,\label{identity-J1}
\\
&& \frac{d\bm{P}}{dt}=\dot{\bm{P}}_{N}+\dot{\bm{P}}_{0.5SO}+\dot{\bm{P}}_{PN}+\dot{\bm{P}}_{1.5SO}~,\label{identity-P1}
\end{eqnarray}
where
\begin{eqnarray}
&&\hskip -0.5cm\dot{E}_{N}=-\frac{8}{15}\frac{G^3M^2\mu^2}{c^5R^4}\Big[12V^2-11\dot{R}^2\Big],\label{dentity-EN}
\end{eqnarray}
\begin{eqnarray}
&&\hskip -2.5cm\dot{E}_{PN}=-\frac{2}{105}\frac{G^3M^2\mu^2}{c^7R^4}\Big[(785-825\eta)V^4-2(1487-1392\eta)V^2\dot{R}^2+3(687-620\eta)\dot{R}^4\nn\\
&&\hskip -2.5cm\hskip 1.4cm-160(17-\eta)\frac{GM}{R}V^2+8(367-15\eta)\frac{GM}{R}\dot{R}^2+16(1-4\eta)\frac{(GM)^2}{R^2}\Big]~,\label{dentity-EPN}
\end{eqnarray}
\begin{eqnarray}
&&\dot{E}_{1.5SO}=-\frac{8}{15}\frac{G^3M\mu^2}{c^8R^5}\Big\{\Big[78\dot{R}^2-8\frac{GM}{R}
-80V^2\Big](\bm{n}\times\bm{V})\cdot\bm{S}\nn\\
&&\hskip 1.4cm+\Big[4\frac{GM}{R}-43V^2+51\dot{R}^2\Big]\frac{\delta M}{M}(\bm{n}\times\bm{V})\cdot\bm{\Delta}\Big\}~,\label{dentity-E1.5SO}
\end{eqnarray}
\begin{eqnarray}
&&\hskip -2.5cm\dot{E}_{2PN}=\!-\frac{2}{2835}\frac{G^3M^2\mu^2}{c^9R^4}\Big\{\Big[18(1692\!-\!5497\eta\!+\!4430\eta^2)V^6\!
-\!24(253\!-\!1026\eta\!+\!56\eta^2)\frac{(GM)^3}{R^3}\nn\\
&&\hskip -1.5cm-\,54(1719-10278\eta+6292\eta^2)V^4\dot{R}^2+108(4987-8513\eta+2165\eta^2)\frac{GM}{R}V^2\dot{R}^2\nn\\
&&\hskip -1.5cm-\,3(106319+9798\eta+5376\eta^2)\frac{(GM)^2}{R^2}\dot{R}^2+54(2018-15207\eta+7572\eta^2)V^2\dot{R}^4\nn\\
&&\hskip -1.5cm-\,12(33510-60971\eta+14290\eta^2)\frac{GM}{R}\dot{R}^4-36(4446-5237\eta+1393\eta^2)\frac{GM}{R}V^4\nn\\
&&\hskip -1.5cm-\,18(2501-20234\eta+8404\eta^2)\dot{R}^6\!+\!(281473+81828\eta+4368\eta^2)\frac{(GM)^2}{R^2}V^2\!\Big]\!\Big\}~,\label{dentity-E2PN}
\end{eqnarray}

\begin{eqnarray}
&&\hskip -2.0cm \dot{E}_{2.5SO}=-\frac{2}{105}\frac{G^3M\mu^2}{c^{10}R^5}\Big\{\Big[(3776+1560\eta)\frac{G^2M^2}{R^2}+(15220-896\eta)\frac{GM}{R}V^2\nn\\
&&\hskip -1.0cm-(12892-2024\eta)\frac{GM}{R}\dot{R}^2-(4828-7240\eta)V^4+(14076-20016\eta)V^2\dot{R}^2\nn\\
&&\hskip -1.0cm-(8976-12576\eta)\dot{R}^4\Big](\bm{n}\times\bm{V})\cdot\bm{S}+\Big[(10774-1988\eta)\frac{GM}{R}V^2\nn\\
&&\hskip -1.0cm-(14654-4796\eta)\frac{GM}{R}\dot{R}^2-(2603-4160\eta)V^4+(9456-14484\eta)V^2\dot{R}^2\nn\\
&&\hskip -1.0cm-(548-952\eta)\frac{G^2M^2}{R^2}-(7941-10704\eta)\dot{R}^4\Big]\frac{\delta M}{M}(\bm{n}\times\bm{V})\cdot\bm{\Delta}\Big\}~,\label{dentity-E2.5SO}
\end{eqnarray}
\begin{eqnarray}
&&\hskip -2.0cm \dot{\bm{J}}_{N}=-\frac{8}{5}\frac{G^2M\mu^2}{c^5R^2}(\bm{n}\!\times\!\bm{V})\Big\{2V^2-3\dot{R}^2+2\frac{GM}{R}\Big\}~,\label{identity-JN}
\end{eqnarray}
\begin{eqnarray}
&&\hskip -2.0cm \dot{\bm{J}}_{PN}=-\frac{2}{105}\frac{G^2M\mu^2}{c^7R^2}(\bm{n}\!\times\!\bm{V})\Big[(307-548\eta)V^4\!
-\!6(74-277\eta)V^2\dot{R}^2\!+\!15(19-72\eta)\dot{R}^4\nn\\
&&\hskip -1.0cm-4(58+95\eta)\frac{GM}{R}V^2\!+\!2(372+197\eta)\frac{GM}{R}\dot{R}^2\!-\!2(745-2\eta)\frac{(GM)^2}{R^2}\Big]~,\label{identity-JPN}
\end{eqnarray}
\begin{eqnarray}
&&\hskip -2.0cm \dot{\bm{J}}_{1.5SO}=\frac{4}{15}\frac{G^2\mu^2}{c^8R^3}(\bm{n}\times\bm{V})\Big\{\Big[163\frac{GM}{R}+111V^2-195\dot{R}^2\Big](\bm{n}\times\bm{V})\cdot\bm{S}\nn\\
&&\hskip -1.0cm+\Big[71\frac{GM}{M}+57V^2-105\dot{R}^2\Big]\frac{\delta M}{M}(\bm{n}\times\bm{V})\cdot\bm{\Delta}\Big\}\nn\\
&&\hskip -1.0cm+\frac{4}{15}\frac{G^2\mu^2}{c^8R^3}V\bm{\lambda}\Big\{\Big[12(\bm{n}\cdot\bm{S})+5\frac{\delta M}{M}(\bm{n}\cdot\bm{\Delta})\Big]\frac{GM}{M}\dot{R}\nn\\
&&\hskip -1.0cm+\Big[50\frac{GM}{R}+71V^2-108\dot{R}^2\Big](\bm{V}\cdot\bm{S})
+\Big[27\frac{GM}{R}+35V^2-54\dot{R}^2\Big]\frac{\delta M}{M}(\bm{V}\cdot\bm{\Delta})\Big\}\nn\\
&&\hskip -1.0cm+\frac{4}{15}\frac{G^2\mu^2}{c^8R^3}\bm{n}\Big\{\Big[41\frac{GM}{R}V^2
-4\frac{(GM)^2}{R^2}-45\frac{GM}{R}\dot{R}^2\Big](\bm{n}\cdot\bm{S})+\Big[165\dot{R}^2-132V^2\nn\\
&&\hskip -1.0cm-54\frac{GM}{R}\Big]\dot{R}(\bm{V}\cdot\bm{S})+\Big[2\frac{(GM)^2}{R^2}+24\frac{GM}{R}V^2-27\frac{GM}{R}\dot{R}^2\Big]\frac{\delta M}{M}(\bm{n}\cdot\bm{\Delta})\nn\\
&&\hskip -1.0cm-\Big[25\frac{GM}{R}+60V^2-75\dot{R}^2\Big]\frac{\delta M}{M}\dot{R}(\bm{V}\cdot\bm{\Delta})\Big\}\nn\\
&&\hskip -1.0cm+\frac{4}{15}\frac{G^2\mu^2}{c^8R^3}\Big\{\bm{S}\Big[4\frac{G^2M^2}{R^2}-91\frac{GM}{R}V^2-71V^4+87\frac{GM}{R}\dot{R}^2
+240V^2\dot{R}^2-165\dot{R}^4\Big]\nn\\
&&\hskip -1.0cm
+\bm{\Delta}\Big[2\frac{G^2M^2}{R^2}+49\frac{GM}{R}V^2+35V^4-45\frac{GM}{R}\dot{R}^2-114V^2\dot{R}^2+75\dot{R}^4\Big]\Big\}~,\label{identity-J1.5SO}
\end{eqnarray}
\begin{eqnarray}
&&\hskip -2.0cm \dot{\bm{J}}_{2PN}=-\frac{1}{2835}\frac{G^2M\mu^2}{c^9R^2}(\bm{n}\!\times\!\bm{V})\Big[(340724\!+\!140922\eta\!+\!2772\eta^2)\frac{(GM)^3}{R^3}\nn\\
&&\hskip -1.0cm-(49140-205380\eta+122220\eta^2)\dot{R}^6\!+\!(23985-111195\eta+116046\eta^2)V^6\nn\\
&&\hskip -1.0cm+(151848-451836\eta+82566\eta^2)\frac{(GM)^2}{R^2}\dot{R}^2\!-\!(200808-372582\eta\nn\\
&&\hskip -1.0cm +87255\eta^2)\frac{GM}{R}\dot{R}^4+(96525-453735\eta+423360\eta^2)V^2\dot{R}^4\!-\!(\!191718\!-\!183222\eta\nn\\
&&\hskip -1.0cm+61704\eta^2)\frac{(GM)^2}{R^2}V^2+(196677-194427\eta+22959\eta^2)\frac{GM}{R}V^2\dot{R}^2-(60642\nn\\
&&\hskip -1.0cm-341631\eta+422199\eta^2)V^4\dot{R}^2+(1485-4419\eta+36198\eta^2)\frac{GM}{R}V^4\Big]~,\label{identity-J2PN}
\end{eqnarray}

\begin{eqnarray}
&&\hskip -2.5cm \dot{\bm{J}}_{2.5SO}=\!-\frac{2}{315}\frac{G^2\mu^2}{c^{10}R^3}(\!\bm{n}\times\bm{V}\!)\!\Big\{\!(\bm{n}\times\bm{V}\!)\!\cdot\!\bm{S}\Big[\!(60751\!+\!4021\eta)\frac{(GM)^2}{R^2}
\!-\!(3996-12501\eta)V^4\nn\\
&&\hskip  -1.5cm+(11340-10710\eta)\dot{R}^4+(5276+18000\eta)\frac{GM}{R}V^2-(8640+4995\eta)V^2\dot{R}^2\nn\\
&&\hskip  -1.5cm-(23799+19353\eta)\frac{GM}{R}\dot{R}^2\Big]+\frac{\delta M}{M}(\bm{n}\times\bm{V})\cdot\bm{\Delta}\Big[(30042+1376\eta)\frac{G^2M^2}{R^2}\nn\\
&&\hskip  -1.5cm+(6030+6041\eta)\frac{GM}{R}V^2-(17850+4164\eta)\frac{GM}{R}\dot{R}^2-(1638-7335\eta)V^4\nn\\
&&\hskip  -1.5cm-(8820+11160\eta)V^2\dot{R}^2+(11970+2205\eta)\dot{R}^4\Big]\Big\}\nn\\
&&\hskip  -1.5cm+\frac{1}{315}\frac{G^2\mu^2}{c^{10}R^3}\bm{n}\Big\{(\bm{n}\cdot\bm{S})\Big[(6360+744\eta)\frac{(GM)^3}{R^3}
-(4013+1388\eta)\frac{GM}{R}V^2\dot{R}^2\nn\\
&&\hskip  -1.5cm-(35306+1742\eta)\frac{(GM)^2}{R^2}V^2\!+\!(61380+4956\eta)\frac{(GM)^2}{R^2}\dot{R}^2\!-\!(3024-5040\eta)V^6\nn\\
&&\hskip  -1.5cm+(12429+5166\eta)\frac{GM}{R}\dot{R}^4+(5645+1254\eta)\frac{GM}{R}V^4+(26460-44100\eta)V^4\dot{R}^2\nn\\
&&\hskip  -1.5cm-(26460-44100\eta)V^2\dot{R}^4\Big]+\frac{\delta M}{M}(\bm{n}\cdot\bm{\Delta})\Big[(30942+5082\eta)\frac{(GM)^2}{R^2}\dot{R}^2\nn\\
&&\hskip  -1.5cm-(17094+2194\eta)\frac{(GM)^2}{R^2}V^2-(720-420\eta)\frac{(GM)^3}{R^3}+(5058-247\eta)\frac{GM}{R}V^4\nn\\
&&\hskip  -1.5cm-(18615+1338\eta)\frac{GM}{R}V^2\dot{R}^2+(7263-9\eta)\frac{GM}{R}\dot{R}^4-(3024-2016\eta)V^6\nn\\
&&\hskip  -1.5cm+(26460-17640\eta)V^4\dot{R}^2-(26460-17640\eta)V^2\dot{R}^4\Big]\nn\\
&&\hskip  -1.5cm+(\bm{V}\cdot\bm{S})\dot{R}\Big[(51034+5742\eta)\frac{GM}{R}V^2
-(84174+8610\eta)\frac{GM}{R}\dot{R}^2\nn\\
&&\hskip  -1.5cm-(17190-8460\eta)V^2\dot{R}^2-(3681-18324\eta)V^4+(37556+6938\eta)\frac{(GM)^2}{R^2}\nn\\
&&\hskip  -1.5cm+(22365-34020\eta)\dot{R}^4\Big]+(\bm{V}\cdot\bm{\Delta})\dot{R}\Big[(14634+3748\eta)\frac{(GM)^2}{R^2}\nn\\
&&\hskip  -1.5cm+(29595+5218\eta)\frac{GM}{R}V^2-(41397+7176\eta)\frac{GM}{R}\dot{R}^2-(5913-7011\eta)V^4\nn\\
&&\hskip  -1.5cm-(5850-6210\eta)V^2\dot{R}^2+(16065-16065\eta)\dot{R}^4\Big]\Big\}\nn\\
&&\hskip  -1.5cm+\frac{1}{315}\frac{G^2\mu^2}{c^{10}R^3}V\bm{\lambda}\Big\{(\bm{n}\cdot\bm{S})\dot{R}\Big[
(3118+10176\eta)\frac{GM}{R}V^2\!-\!(27946+1222\eta)\frac{(GM)^2}{R^2}\nn\\
&&\hskip  -1.5cm+(1902-8124\eta)\frac{GM}{R}\dot{R}^2+(3024-5040\eta)V^4
-(26460-44100\eta)V^2\dot{R}^2\nn\\
&&\hskip  -1.5cm+(26460-44100\eta)\dot{R}^4\Big]-(\bm{V}\cdot\bm{S})\Big[(20890+10614\eta)\frac{GM}{R}V^2\nn\\
&&\hskip  -1.5cm+(33212+4178\eta)\frac{(GM)^2}{R^2}-(47262+10002\eta)\frac{GM}{R}\dot{R}^2-(2247-10974\eta)V^4\nn\\
&&\hskip  -1.5cm-(22752-20448\eta)V^2\dot{R}^2+(25965-38970\eta)\dot{R}^4\Big]\nn\\
&&\hskip  -1.5cm+\frac{\delta M}{M}\dot{R}(\bm{n}\cdot\bm{\Delta})\Big[(5313-2496\eta)\frac{GM}{R}\dot{R}^2-(26460-17640\eta)V^2\dot{R}^2\nn\\
&&\hskip  -1.5cm+(885+3847\eta)\frac{GM}{R}V^2+(26460-17640\eta)\dot{R}^4-(13896+1136\eta)\frac{(GM)^2}{R^2}\nn\\
&&\hskip  -1.5cm+(3024-2016\eta)V^4\Big]-\frac{\delta M}{M}(\bm{V}\cdot\bm{\Delta})\Big[(10995+6622\eta)\frac{GM}{R}V^2\nn\\
&&\hskip  -1.5cm+(14970+1552\eta)\frac{(GM)^2}{R^2}-(20661+6012\eta)\frac{GM}{R}\dot{R}^2\nn\\
&&\hskip  -1.5cm-(663-6147\eta)V^4-(19908-4122\eta)V^2\dot{R}^2+(24345-13545\eta)\dot{R}^4\Big]\Big\}\nn\\
&&\hskip  -1.5cm+\frac{1}{315}\frac{G^2\mu^2}{c^{10}R^3}\Big\{\bm{S}\Big[(14731+14916\eta)\frac{GM}{R}V^4-(75236+6410\eta)\frac{(GM)^2}{R^2}\dot{R}^2\nn\\
&&\hskip  -1.5cm+\!(\!72764+1658\eta\!)\frac{(GM)^2}{R^2}V^2\!-\!(80308+28326\eta)\frac{GM}{R}V^2\dot{R}^2\!-\!(6360+744\eta)\frac{(GM)^3}{R^3}\nn\\
&&\hskip  -1.5cm+(73329+16746\eta)\frac{GM}{R}\dot{R}^4-(2247-10974\eta)V^6-(19071-2124\eta)V^4\dot{R}^2\nn\\
&&\hskip  -1.5cm+(43155-47430\eta)V^2\dot{R}^4-(22365-34020\eta)\dot{R}^6\Big]
+\frac{\delta M}{M}\bm{\Delta}\Big[(720-420\eta)\frac{(GM)^3}{R^3}\nn\\
&&\hskip  -1.5cm+\!(\!36228+2050\eta\!)\frac{(GM)^2}{R^2}V^2\!-\!(35844+5998\eta)\frac{(GM)^2}{R^2}\dot{R}^2\!+\!(\!34089\!+\!11577\eta\!)\frac{GM}{R}\dot{R}^4\nn\\
&&\hskip  -1.5cm+(6915+8983\eta)\frac{GM}{R}V^4-(38772+17776\eta)\frac{GM}{R}V^2\dot{R}^2-(663-6147\eta)V^6\nn\\
&&\hskip  -1.5cm-(13995+2889\eta)V^4\dot{R}^2+(30195-19755\eta)V^2\dot{R}^4-(16065-16065\eta)\dot{R}^6\Big]\Big\},~\label{identity-J2.5PN}
\end{eqnarray}
\begin{eqnarray}
&&\dot{\bm{P}}_{N}=-\frac{8}{105}\frac{G^3M^2\mu^2}{c^7R^4}(1-4\eta)^\frac{1}{2}
\Big\{{V\bm{\lambda}}\Big[38\dot{R}^2-50V^2-8\frac{GM}{R}\Big]\nn \\
&&\hskip 1cm +\dot{R}\bm{n}\Big[55V^2+12\frac{GM}{R}-45\dot{R}^2\Big]\Big\}~,\label{identity-PN}\\
\nn\\ \nn
\end{eqnarray}
\begin{eqnarray}
&&\dot{\bm{P}}_{0.5SO}=-\frac{8}{15}\frac{G^3M\mu^2}{c^8R^5}\Big\{4\dot{R}(\bm{V}\times\bm{\Delta})-2V^2(\bm{n}\times\bm{\Delta})\nn\\
&&\hskip 1.0cm -(\bm{n}\times\bm{V})\Big[3\dot{R}(\bm{n}\cdot\bm{\Delta})+2(\bm{V\cdot\bm{\Delta}})\Big]\Big\}~,\label{identity-P0.5SO}
\end{eqnarray}
\begin{eqnarray}
&&\hskip -2cm \dot{\bm{P}}_{PN}=-\frac{1}{945}\frac{G^3M^2\mu^2}{c^9R^4}(1-4\eta)^\frac{1}{2}\nn\\
&&\hskip -1cm\times \Big\{{V\bm{\lambda}}\Big[32(189+17\eta)\frac{(GM)^2}{R^2}-12(2663-1394\eta)\dot{R}^4,36(907-162\eta)\frac{GM}{R}V^2\nn\\
&&\hskip -1cm+120(392-257\eta)V^2\dot{R}^2-444(25-28\eta)V^4-12(2699+10\eta)\frac{GM}{R}\dot{R}^2\Big]\nn\\
&&\hskip -1cm+\dot{R}\bm{n}\Big[4(12301\!-\!1168\eta)\frac{GM}{R}\dot{R}^2\!+\!24(851\!-\!779\eta)V^4\!-\!24(2834\!-\!1877\eta)V^2\dot{R}^2\nn\\
&&\hskip -1cm\!-12(590\!-\!4\eta)\frac{(GM)^2}{R^2}\!+\!24(1843\!-\!1036\eta)\dot{R}^4\!-\!12(4385\!-\!956\eta)\frac{GM}{R}V^2\Big]\Big\}~,\label{identity-PPN}
\end{eqnarray}
\begin{eqnarray}
&&\hskip -2.5cm\dot{\bm{P}}_{1.5SO}=\frac{4}{945}\frac{G^3M\mu^2}{c^{10}R^5}(\bm{n}\times\bm{V})\Big\{\Big[21852\dot{R}^2-19536V^2
+2166\frac{GM}{R}\Big]\dot{R}\frac{\delta M}{M}(\bm{n}\cdot\bm{S})\nn\\
&&\hskip -1.5cm+\Big[-272\frac{GM}{R}-4857\dot{R}^2+3314V^2\Big]\dot{R}\frac{\delta M}{M}(\bm{V}\cdot\bm{S})
+\Big[-(4902+2919\eta)\frac{GM}{R}\nn \\
&&\hskip -1.5cm+(6264-37062\eta)\dot{R}^2
-(5172-35448\eta)V^2\!\Big]\dot{R}(\bm{n}\cdot\bm{\Delta})
\!+\!\Big[\!-(572-647\eta)\frac{GM}{R}\nn \\
&&\hskip -1.5cm-(4281-8556\eta)\dot{R}^2+(2024-5621\eta)V^2\Big](\bm{V}\cdot\bm{\Delta})\Big\}\nn\\
&&\hskip -1.5cm+\frac{4}{945}\frac{G^3M\mu^2}{c^{10}R^5}\bm{n}\dot{R}\Big\{
\Big[3180\frac{GM}{R}-17964\dot{R}^2+17592V^2\Big]\frac{\delta M}{M}(\bm{n}\times\bm{V})\cdot\bm{S}\Big\}\nn\\
&&\hskip -1.5cm+\Big[\!(1509-5451\eta)\frac{GM}{R}+(9492-35934\eta)V^2-(9396-35514\eta)\dot{R}^2\Big](\bm{n}\times\bm{V})\cdot\bm{\Delta}\nn \\
&&\hskip -1.5cm+\frac{4}{945}\frac{G^3M\mu^2}{c^{10}R^5}V\bm{\lambda}\Big\{-\Big[7985V^2+2176\frac{GM}{R}-8043\dot{R}^2\Big]\frac{\delta M}{M}(\bm{n}\times\bm{V})\cdot\bm{S}\nn\\
&&\hskip -1.5cm+\Big[(3822-14361\eta)\dot{R}^2-(3968-15017\eta)V^2-(697-2503\eta)\frac{GM}{R}\Big](\bm{n}\times\bm{V})\cdot\bm{\Delta}\Big\}\nn\\
&&\hskip -1.5cm+\frac{4}{945}\frac{G^3M\mu^2}{c^{10}R^5}\Big\{-\frac{\delta M}{M}\dot{R}\Big[608\frac{GM}{R}-5709\dot{R}^2+5431V^2\Big](\bm{V}\times\bm{S})\nn\\
&&\hskip -1.5cm-\Big[(3585+4506\eta)\frac{GM}{R}\dot{R}^2+(10287-9810\eta)\dot{R}^4+(1417-3484\eta)\frac{GM}{R}V^2\nn \\
&&\hskip -1.5cm+(10287-9810\eta)V^2\dot{R}^2+(1322-968\eta)V^4\Big](\bm{n}\times\bm{\Delta})\nn\\
&&\hskip -1.5cm -\dot{R}\Big[(5677-15868\eta)V^2-(8355-18048\eta)\dot{R}^2-(6478-184\eta)\frac{GM}{R}\Big](\bm{V}\times\bm{\Delta})\nn\\
&&\hskip -1.5cm +\Big[1812\frac{GM}{R}\dot{R}^2-6300\dot{R}^4-664\frac{GM}{R}V^2+7491V^2\dot{R}^2-1469V^4\nn \\
&&\hskip -1.5cm+72\frac{(GM)^2}{R^2}\Big]\frac{\delta M}{M}(\bm{n}\times\bm{S})\Big\}~.\label{identity-P1.5SO}
\end{eqnarray}

All the calculations have been done with the help of {\tt MATHEMATICA}.

Among the above results, the loss rate of the system's energy given in Eqs.~(\ref{dentity-EN})-(\ref{dentity-E2PN}) has been achieved in Ref.~\cite{WillWiseman1996}. The Newtonian and 1PN contributions to the loss rate of the system's angular momentum, see Eqs.~(\ref{identity-JN}) and (\ref{identity-JPN}), as well as the Newtonian contribution to the linear momentum loss rate, see Eq.~(\ref{identity-PN}), have been obtained in Ref.~\cite{JunkerSchafer1992}. The 2PN contributions to the loss rate of the system's angular momentum and the 1PN contributions the linear momentum loss rate, see Eqs.~(\ref{identity-J2PN}) and ~(\ref{identity-PPN}), have been obtained in Ref~\cite{Arun2009,Racine2009}.
The 1.5PN SO contributions to the loss rates of the energy and angular momentum, see Eqs.~(\ref{dentity-E1.5SO}) and (\ref{identity-J1.5SO}), as well as the 0.5PN SO contribution to the linear momentum loss rate, see (\ref{identity-P0.5SO}), have been achieved in Ref~\cite{Kidder1995}. Here we include them for the completeness.

The core results of this work are the loss rates of the system's energy and angular momentum induced by the 2.5PN SO coupling effect, see Eqs.~(\ref{dentity-E2.5SO}) and (\ref{identity-J2.5PN}), and the linear momentum's loss rate induced by the 1.5PN SO coupling effect, see Eq.~(\ref{identity-P1.5SO}). Notice that all these loss can be called as the next-to-leading spin-orbit coupling effects, since the leading spin-orbit coupling effects to the system's energy and angular momentum are the 1.5PN SO coupling contribution, while the leading spin-orbit coupling effect to the system's linear momentum is the 0.5PN SO coupling contribution.

\section{Loss rates of gravitational radiation in the case of circular orbit}\label{sec5}
When the binary systems loss their orbital energy, angular momentum and linear momentum due to the gravitational-wave radiation, their orbit will shrink and their eccentricity will decrease. In the final stage of binary inspiral, their orbit can be approximated as circular ones~\cite{LincolnWill1990,Kidder1995}. In this case, we have $\dot{R}\!=\!0$ and $\bm{R}\!\times\!\bm{V}\!=\!RV\bm{l}$.
Following Blanchet~\cite{Blanchet2002}, we introduce the PN parameter $x=\big(\frac{GM\omega}{c^3}\big)^{\frac{2}{3}}$ with
\begin{eqnarray}
&&\hskip -2cm\omega^2= \frac{GM}{R^3}\Big\{1-\frac{GM}{c^2R}(3-\eta)+\Big(\frac{GM}{c^2R}\Big)^2\Big(6+\frac{41}{4}\eta+\eta^2\Big)
\!-\!\Big(\frac{GM}{c^2R}\Big)^{\frac{3}{2}}\Big(5\frac{S_{\bm{l}}}{M^2}+3\frac{\delta M }{M}\frac{\Delta_{\bm{l}}}{M^2}\Big)\nn\\
&&\hskip -1cm  +\Big(\frac{GM}{c^2R}\Big)^{\frac{5}{2}}\Big[\Big(\frac{45}{2}-\frac{27}{2}\eta\Big)\frac{S_{\bm{l}}}{M^2}+\Big(\frac{27}{2}-\frac{13}{2}\eta\Big)\frac{\delta M }{M}\frac{\Delta_{\bm{l}}}{M^2}\Big]\Big\}~,\label{Circular orbits-frequency}
\end{eqnarray}
where $\omega$ is the orbital frequency, then we can obtain:
\begin{eqnarray}
&&\hskip -2.5cm\frac{dE}{dt}=-\frac{32}{5}\frac{c^5}{G}x^5 \eta^2\Big\{1-x\Big(\frac{1247}{336}+\frac{35}{12}\eta\Big)
-x^2\Big(\frac{44711}{9072}-\frac{9271}{504}\eta-\frac{65}{18}\eta^2\Big)\nn\\
&&\hskip -1.5cm -x^{\frac{3}{2}}\frac{1}{G}\Big(\!4\frac{S_{\bm{l}}}{M^2}+\frac{5}{4}\frac{\delta M}{M}\frac{\Delta_{\bm{l}}}{M^2}\!\Big)
-x^{\frac{5}{2}}\frac{1}{G}\Big[\!\Big(\frac{9}{2}-\frac{272}{9}\eta\Big)\frac{S_{\bm{l}}}{M^2}+\frac{\delta M}{M}\Big(\!\frac{13}{16}-\frac{43}{4}\eta\!\Big)\frac{\Delta_{\bm{l}}}{M^2}\!\Big]\Big\}~,\label{Circular orbits-E111}
\end{eqnarray}
\begin{eqnarray}
&&\hskip -2.5cm\frac{d\bm{J}}{dt}=-\frac{32}{5}Mc^2x^{\frac{7}{2}}\eta^2\Big\{1-x\Big(\frac{1247}{336}+\frac{35\eta}{12}\Big)
-x^2\Big(\frac{44711}{9072}-\frac{9271}{504}\eta-\frac{65}{18}\eta^2\Big)\nn \\
&&\hskip -1.5cm -x^{\frac{3}{2}}\frac{1}{G}\Big(\!4\frac{S_{\bm{l}}}{M^2}+\frac{\delta M}{M}\frac{5}{4}\frac{\Delta_{\bm{l}}}{M^2}\!\Big)
-x^{\frac{5}{2}}\frac{1}{G}\Big[\Big(\frac{9}{2}-\frac{272}{9}\eta\Big)\frac{S_{\bm{l}}}{M^2}+\frac{\delta M}{M}\Big(\frac{13}{16}-\frac{43}{4}\eta\Big)\frac{\Delta_{\bm{l}}}{M^2}\Big]\Big\}\bm{l}\nn \\
&&\hskip -1.5cm -\frac{32}{5}Mc^2x^{\frac{7}{2}}\eta^2\Big\{x^{\frac{3}{2}}\Big(\frac{121}{24}\frac{S_{\bm{n}}}{GM^2}+\frac{5}{2}\frac{\delta M}{M}\frac{\Delta_{\bm{n}}}{GM^2}\Big)-x^{\frac{5}{2}}\Big[\Big(\frac{2387}{96}+\frac{2057}{126}\eta\Big)\frac{S_{\bm{n}}}{GM^2}\nn\\
&&\hskip -1.5cm +\Big(\frac{545}{42}+\frac{5725}{672}\eta\Big)\frac{\delta M}{M}\frac{\Delta_{\bm{n}}}{GM^2}\Big]\Big\}\bm{n}-\frac{32}{5}Mc^2x^{\frac{7}{2}}\eta^2\Big\{x^{\frac{3}{2}}\Big(\frac{37}{24}\frac{S_{\bm{\lambda}}}{GM^2}
+\frac{\delta M}{M}\frac{\Delta_{\bm{\lambda}}}{GM^2}\Big)\nn\\
&&\hskip -1.5cm -x^{\frac{5}{2}}\Big[\Big(\frac{2425}{224}+\frac{1387}{1008}\eta\Big)\frac{S_{\bm{\lambda}}}{GM^2}+\Big(\frac{2227}{336}+\frac{439}{224}\eta\Big)\frac{\delta M}{M}\frac{\Delta_{\bm{\lambda}}}{GM^2}\Big]\Big\}\bm{\lambda}~,\label{Circular orbits-J111}
\end{eqnarray}
\begin{eqnarray}
&&\hskip -2.5cm\frac{d\bm{P}}{dt}=\frac{464}{105}\frac{c^4}{G}x^{\frac{11}{2}}\eta^2(1-4\eta)^{\frac{1}{2}}\Big\{1-x\Big(\frac{452}{87}+\frac{1139}{522}\eta
\Big)-x^{\frac{1}{2}}\frac{7}{29}\frac{\Delta_{\bm{l}}}{GM^2}\nn\\
&&\hskip -1.5cm -x^{\frac{3}{2}}\Big[\frac{470}{87}\frac{\delta M}{M}\frac{S_{\bm{l}}}{GM^2}
+\Big(\frac{67}{58}-\frac{206}{29}\eta\Big)\frac{\Delta_{\bm{l}}}{GM^2}\Big]\Big\}\bm{\lambda }\nn\\
&&\hskip -1.5cm +\frac{464}{105}\frac{c^4}{G}x^{\frac{11}{2}}\eta^2(1-4\eta)^{\frac{1}{2}}\Big\{x^{\frac{1}{2}}\frac{14}{29}\frac{\Delta_{\bm{\lambda}}}{GM^2}+x^{\frac{3}{2}}\Big[\frac{109}{116}\frac{\delta M}{M}\frac{S_{\bm{\lambda}}}{GM^2}\nn \\
&&\hskip -1.5cm+\Big(\frac{25}{116}-\frac{57}{58}\eta\Big)\frac{\Delta_{\bm{\lambda}}}{GM^2}\Big]\Big\}\bm{l}~.\label{Circular orbits-P111}
\end{eqnarray}

here $S_{\bm{l}}=\bm{l}\cdot\bm{S}$, $S_{\bm{n}}=\bm{n}\cdot\bm{S}$, $S_{\bm{\lambda}}=\frac{\bm{V}\cdot\bm{S}}{V}$, and $\Delta_{\bm{l}}=\bm{l}\cdot\bm{\Delta}$, $\Delta_{\bm{n}}=\bm{n}\cdot\bm{\Delta}$, $\Delta_{\bm{\lambda}}=\frac{\bm{V}\cdot\bm{\Delta}}{V}$. Eqs.~(\ref{Circular orbits-E111}) have been obtained in ref~\cite{Marsat2014}. And we have also found that the loss of angular momentum and energy in the $\bm{l}$ direction are equal, which satisfies the formula given by Ref~\cite{Racine2009}:
{\small\begin{eqnarray}
\dot{E}=\omega \,\bm{\dot{J}}\cdot{\bm{l}}~.
\end{eqnarray}}
\section{Summary}\label{sec6}

Based on the spin vector defined by Boh\'{e} et al. and under the T condition, we have calculated the loss rates for the binary systems' energy, angular and linear momentum induced by the next-to-leading spin-orbit coupling effects in the case of general orbit. For comparison, we have also adopted the spin vector defined by Faye, Blanchet and Buonanno to calculate the angular momentum's loss rate induced by the 2.5PN spin-orbit coupling effect. For the case of circular orbit, we formulate these gravitational losses in terms of the orbital frequency. The achieved results are useful in determining the time change of the orbital parameters or the general motion when the spin vector defined by Boh\'{e} et al. is adopted.

\appendix
\section{The STF tensors used in the derivations}
For the readers' convenience, we give the STF tensors used in the derivations for the compact binary systems' gravitational losses.
\begin{eqnarray}
&&\hskip -1.0cm  A_{<i}B_{j>}=\frac{1}{2}(A_iB_j+B_iA_j)-\frac{1}{3}\delta_{ij}A_aB^a~,\label{Aij}
\end{eqnarray}
\begin{eqnarray}
&&\hskip -2.0cm  A_{<i}B_jC_{k>}=\frac{1}{6}(A_iB_jC_k+A_iB_kC_j+A_jB_iC_k+A_jB_kC_i+A_kB_iC_j+A_kB_jC_i)\nn\\
&&\hskip 0cm-\,\frac{1}{15}[(\delta_{ij}C_k+\delta_{jk}C_i+\delta_{ik}C_j)A_aB^a+(\delta_{ij}B_k+\delta_{jk}B_i+\delta_{ik}B_j)A_aC^a\nn \\
&&\hskip -0cm+(\delta_{ij}A_k+\delta_{jk}A_i+\delta_{ik}A_j)B_aC^a]~,\label{Aijk}
\end{eqnarray}
\begin{eqnarray}
&&\hskip -1.5cm  A_{<ijkl>}=A_iA_jA_kA_l-\frac{1}{7}A_aA^a(\delta_{ij}A_kA_l+\delta_{ik}A_iA_l+\delta_{il}A_jA_k+\delta_{jk}A_iA_l+\delta_{jl}A_iA_k\nn\\
&&\hskip 0cm+\,\delta_{kl}A_iA_j)+\frac{1}{35}A_aA^aA_bA^b(\delta_{ij}\delta_{kl}+\delta_{ik}\delta_{jl}+\delta_{il}\delta_{jk})~,\label{Aijkl}
\end{eqnarray}
\begin{eqnarray}
&&\hskip -0.3cm  \epsilon_{ab<i}A_{j>}B_aC_b=\frac{1}{2}(\epsilon_{ab<i}A_{j>}+\epsilon_{ab<j}A_{i>})B_aC_b-\frac{1}{3}\delta_{ij}\epsilon_{abk}A_{k}B_aC_b~,\label{Aij}
\end{eqnarray}
\begin{eqnarray}
&&\hskip -2.0cm  \epsilon_{ab<i}A_jB_{k>}C_aD_b=\Big\{\frac{1}{6}(\epsilon_{abi}A_jB_k+\epsilon_{abi}A_kB_j+\epsilon_{abj}A_iB_k+\epsilon_{abj}A_kB_i+\epsilon_{abk}A_iB_j\nn\\
&&\hskip -1cm\hskip 2.0cm+\,\epsilon_{abk}A_jB_i)-\frac{1}{15}[(\delta_{ij}B_k+\delta_{ik}B_j+\delta_{jk}B_i)\epsilon_{abl}A_l+(\delta_{ij}A_k+\delta_{ik}A_j\nn\\
&&\hskip -1cm\hskip 2.0cm+\delta_{jk}A_i)\epsilon_{abl}B_l+(\delta_{ij}\epsilon_{abk}+\delta_{ik}\epsilon_{abj}+\delta_{jk}\epsilon_{abi})A_lB^l]\Big\}C_aD_b~.\label{Aijk}
\end{eqnarray}

\section{The angular momentum loss induced by the 2.5PN SO effect in terms of the spin vector defined by Faye, Blanchet and Buonanno}
Here we consider the spin vector $\tilde{S}_A^{\mu}$ defined by Faye, Blanchet and Buonanno~\cite{FayeBlanchetBuonanno2006}.
We assume the spins of the binary are $\tilde{\bm{S}}_1$ and $\tilde{\bm{S}}_2$. The total spin vector can be written as $\tilde{\bm{S}}=\tilde{\bm{S}_1}+\tilde{\bm{S}_2}$, $\tilde{\bm{\Delta}}=M\Big(\frac{\tilde{\bm{S}}_2}{M_2}-\frac{\tilde{\bm{S}}_1}{M_1}\Big)$. Then we calculate the angular momentum loss induced by the 2.5PN SO effect for general orbit, which can be written as

\begin{eqnarray}
&&\hskip -2.5cm \dot{\bm{J}}_{2.5SO}=\frac{1}{315}\frac{G^2\mu^2}{c^{10}R^3}(\bm{n}\times\bm{V})\Big\{(\bm{n}\times\bm{V})\cdot\tilde{\bm{S}}\Big[(1746-10800\eta)V^4
\nn\\
&&\hskip -1.5cm -(106022+24242\eta)\frac{(GM)^2}{R^2}-(87570-163170\eta)\dot{R}^4-(20458-7326\eta)\frac{GM}{R}V^2\nn\\
&&\hskip -1.5cm +(76320-121770\eta)V^2\dot{R}^2+(60612-33348\eta)\frac{GM}{R}\dot{R}^2\Big]\nn\\
&&\hskip -1.5cm+\frac{\delta M}{M}(\bm{n}\times\bm{V})\cdot\tilde{\bm{\Delta}}\Big[(40326-39300\eta)\frac{GM}{R}\dot{R}^2-(16194-15062\eta)\frac{GM}{R}V^2\nn\\
&&\hskip -1.5cm -(48876+13480\eta)\frac{G^2M^2}{R^2}+(1674-6822\eta)V^4+(30240-49320\eta)V^2\dot{R}^2\nn\\
&&\hskip -1.5cm -(34650-71190\eta)\dot{R}^4\Big]\Big\}\nn\\
&&\hskip -1.5cm +\frac{1}{315}\frac{G^2\mu^2}{c^{10}R^3}\bm{n}\Big\{(\bm{n}\cdot\tilde{\bm{S}})\Big[(5064+3792\eta)\frac{(GM)^3}{R^3}
-(65118-101184\eta)\frac{GM}{R}V^2\dot{R}^2\nn\\
&&\hskip -1.5cm -(36782-12850\eta)\frac{(GM)^2}{R^2}V^2+(62988-5160\eta)\frac{(GM)^2}{R^2}\dot{R}^2-(4032-8064\eta)V^6\nn\\
&&\hskip -1.5cm +(48123-101916\eta)\frac{GM}{R}\dot{R}^4\!+\!(13769-16314\eta)\frac{GM}{R}V^4\!+\!(35280\!-\!70560\eta)V^4\dot{R}^2\nn\\
&&\hskip -1.5cm -(35280-70560\eta)V^2\dot{R}^4\Big]+\frac{\delta M}{M}(\bm{n}\cdot\tilde{\bm{\Delta}})\Big[(25290+5346\eta)\frac{(GM)^2}{R^2}\dot{R}^2\nn\\
&&\hskip -1.5cm -(15270-3170\eta)\frac{(GM)^2}{R^2}V^2-(2280-2940\eta)\frac{(GM)^3}{R^3}+(8034-7567\eta)\frac{GM}{R}V^4\nn\\
&&\hskip -1.5cm -(31143-44670\eta)\frac{GM}{R}V^2\dot{R}^2+(18009-46449\eta)\frac{GM}{R}\dot{R}^4-(4032-4032\eta)V^6\nn\\
&&\hskip -1.5cm +(35280-35280\eta)V^4\dot{R}^2-(35280-35280\eta)V^2\dot{R}^4\Big]\nn\\
&&\hskip -1.5cm +(\bm{V}\cdot\tilde{\bm{S}})\dot{R}\Big[(46570-4842\eta)\frac{GM}{R}V^2
-(78072-8868\eta)\frac{GM}{R}\dot{R}^2\nn\\
&&\hskip -1.5cm +(57510-219420\eta)V^2\dot{R}^2-(26289-89172\eta)V^4+(36434-1780\eta)\frac{(GM)^2}{R^2}\nn\\
&&\hskip -1.5cm -(31185-126630\eta)\dot{R}^4\Big]+\frac{\delta M}{M}(\bm{V}\cdot\tilde{\bm{\Delta}})\dot{R}\Big[(12192+760\eta)\frac{(GM)^2}{R^2}\nn\\
&&\hskip -1.5cm +(24123+610\eta)\frac{GM}{R}V^2-(30939+1560\eta)\frac{GM}{R}\dot{R}^2-(14265-40995\eta)V^4\nn\\
&&\hskip -1.5cm +(15390-93510\eta)V^2\dot{R}^2+(4095+49455\eta)\dot{R}^4\Big]\Big\}\nn\\
&&\hskip -1.5cm +\frac{1}{315}\frac{G^2\mu^2}{c^{10}R^3}V\bm{\lambda}\Big\{(\bm{n}\cdot\tilde{\bm{S}})\dot{R}\Big[
(25576-64002\eta)\frac{GM}{R}V^2-(29974-1166\eta)\frac{(GM)^2}{R^2}\nn\\
&&\hskip -1.5cm -(23046-80328\eta)\frac{GM}{R}\dot{R}^2+(4032-8064\eta)V^4
-(35280-70560\eta)V^2\dot{R}^2\nn\\
&&\hskip -1.5cm +(35280-70560\eta)\dot{R}^4\Big]-(\bm{V}\cdot\tilde{\bm{S}})\Big[(18730+504\eta)\frac{GM}{R}V^2\nn\\
&&\hskip -1.5cm +(33386-7516\eta)\frac{(GM)^2}{R^2}-(47052-18642\eta)\frac{GM}{R}\dot{R}^2-(8637-30144\eta)V^4\nn\\
&&\hskip -1.5cm +(15066-89982\eta)V^2\dot{R}^2-(6345-54180\eta)\dot{R}^4\Big]\nn \\
&&\hskip -1.5cm -\frac{\delta M}{M}\dot{R}(\bm{n}\cdot\tilde{\bm{\Delta}})\Big[(6567-41892\eta)\frac{GM}{R}\dot{R}^2+(35280-35280\eta)V^2\dot{R}^2\nn\\
&&\hskip -1.5cm -(11859-33338\eta)\frac{GM}{R}V^2-(35280-35280)\dot{R}^4+(12492+1772\eta)\frac{(GM)^2}{R^2}\nn\\
&&\hskip -1.5cm -(4032-3032\eta)V^4\Big]+\frac{\delta M}{M}(\bm{V}\cdot\tilde{\bm{\Delta}})\Big[(18075-9882\eta)\frac{GM}{R}\dot{R}^2\nn\\
&&\hskip -1.5cm -(14088-3956\eta)\frac{(GM)^2}{R^2}-(8835+400\eta)\frac{GM}{R}V^2+(3885-15759\eta)V^4\nn\\
&&\hskip -1.5cm +(4462+45918\eta)V^2\dot{R}^2-(13815+25515\eta)\dot{R}^4\Big]\Big\}\nn \\
&&\hskip -1.5cm +\frac{1}{315}\frac{G^2\mu^2}{c^{10}R^3}\Big\{\tilde{\bm{S}}\Big[(3901+31614\eta)\frac{GM}{R}V^4-(73736-5218\eta)\frac{(GM)^2}{R^2}\dot{R}^2\nn\\
&&\hskip -1.5cm -(5064+3792\eta)\frac{(GM)^3}{R^3}+(54591+37932\eta)\frac{GM}{R}\dot{R}^4-(8637-30144\eta)V^6\nn\\
&&\hskip -1.5cm-(54796+53706\eta)\frac{GM}{R}V^2\dot{R}^2 +(74456-19882\eta)\frac{(GM)^2}{R^2}V^2\nn\\
&&\hskip -1.5cm +(41355-179154\eta)V^4\dot{R}^2-(63855-273600\eta)V^2\dot{R}^4+(31385-126630\eta)\dot{R}^6\Big]\nn\\
&&\hskip -1.5cm +\frac{\delta M}{M}\tilde{\bm{\Delta}}\Big[(2280-2940\eta)\frac{(GM)^3}{R^3}+(33564-6890\eta)\frac{(GM)^2}{R^2}V^2\nn\\
&&\hskip -1.5cm -(29196+4570\eta)\frac{(GM)^2}{R^2}\dot{R}^2+(22857+20109\eta)\frac{GM}{R}\dot{R}^4\nn\\
&&\hskip -1.5cm +(1233+15751\eta)\frac{GM}{R}V^4-(26724+23836\eta)\frac{GM}{R}V^2\dot{R}^2-(3885-15759\eta)V^6\nn\\
&&\hskip -1.5cm +(9603-86913\eta)V^4\dot{R}^2-(1575-119025\eta)V^2\dot{R}^4-(4095+49455\eta)\dot{R}^6\Big]\Big\}~.\label{identity-J2.5SO06}
\end{eqnarray}
For the case of circular orbit, we have
\begin{eqnarray}
&&\dot{\bm{J}}_{2.5SO}=\frac{32}{5}\frac{Mc^2\eta^2}{GM^2}\tilde{x}^{6}\Big\{\Big[(\frac{95}{28}+\frac{239}{63}\eta\Big)\tilde{S}_{\bm{l}}+\frac{\delta M}{M}\Big(\frac{31}{16}-\frac{109}{28}\eta\Big)\tilde{\Delta}_{\bm{l}}\Big]\bm{l}\nn \\
&&\hskip 2.0cm +\Big[\Big(\frac{4471}{224}+\frac{5911}{252}\eta\Big)\tilde{S}_{\bm{n}}+\Big(\frac{383}{42}+\frac{1175}{96}\eta\Big)\frac{\delta M}{M}\tilde{\Delta}_{\bm{n}}\Big]\bm{n}\nn\\
&&\hskip 2.0cm +\Big[\Big(\frac{5323}{672}+\frac{149}{18}\eta\Big)\tilde{S}_{\bm{\lambda}}+\Big(\frac{229}{48}+\frac{1221}{224}\eta\Big)\frac{\delta M}{M}\tilde{\Delta}_{\bm{\lambda}}\Big]\bm{\lambda}\Big\}~.\label{Circular orbits-J11106}
\end{eqnarray}
where the parameter $\tilde{x}=\big(\frac{GM\tilde{\omega}}{c^3}\big)^{\frac{2}{3}}$ can be found in reference~\cite{BlanchetBuonanno2006},
$\tilde{S}_{\bm{l}}=\bm{l}\cdot\tilde{\bm{S}}$, $\tilde{S}_{\bm{n}}=\bm{n}\cdot\tilde{\bm{S}}$, $\tilde{S}_{\bm{\lambda}}=\frac{\bm{V}\cdot\tilde{\bm{S}}}{V}$, and $\tilde{\Delta}_{\bm{l}}=\bm{l}\cdot\bm{\tilde{\Delta}}$, $\tilde{\Delta}_{\bm{n}}=\bm{n}\cdot\bm{\tilde{\Delta}}$, $\tilde{\Delta}_{\bm{\lambda}}=\frac{\bm{V}\cdot\bm{\tilde{\Delta}}}{V}$.

\section*{Acknowledgement}
This study was supported in part by the National Natural Science Foundation of China (Nos. 11973025, U1931204 and 12147208).



\section*{References}
This study was supported in part by the National Natural Science Foundation of China (Nos. 11973025, U1931204 and 12147208).



\end{document}